\documentclass[10pt,psfig,graphicx,twocolumn]{aa}

\usepackage{epsfig, txfonts, natbib}
\bibpunct{(}{)}{;}{a}{}{,}

\begin{document}

\title{Branes in Supernova Shells}

\author{A. S. Popescu}
\titlerunning{Branes in Supernova Shells}

   \institute{Astronomical Institute of Romanian Academy of Science, Str. Cutitul de Argint 5, RO-040557 
Bucharest, Romania\\
              \email{sabinp@aira.astro.ro}}

   \date{Received/Accepted}

   \abstract{

This paper is firstly intended to review shortly the most recent developments and ideas resulting from the necessity of having a scale at which gravitation to unify the other fundamental forces. With the declared intention of predicting an {\it in situ} distinct possibility for Kaluza-Klein gravitons formation in the supernova shell we are using state of the art simulations \citep{langer,ud1,ud2,ud3} for massive stellar winds to infer that the supernova shock which will hit such winds will meet a non-isotropic and non-homogeneous matter distribution with a very distinct geometry. By linking this to the mechanism of particle shock acceleration at cosmic ray energies (Fermi acceleration) and the related spallation in the wind shell, the result is the creation of proper conditions for Kaluza-Klein gravitons formation in the supernova shell from neutrino secondary particles (cosmic ray spallation products) interacting with $\sim 10^{18}$ eV cosmic rays.

   }

   \maketitle
   
\section{Introduction: Brane-World Scenarios}

The main motivation for considering the spacetime as multi-dimensional, by introducing extra-dimensions, comes from the need to provide an unified theory in which two separate scales, the electroweak $M_{W}\sim 100$ GeV and the Planck scale $M_{P}\sim 10^{19}$ GeV can coexist (hierarchy problem). The string theory and M-theory are able to incorporate gravity in a reliable manner. Recently a new multi-dimensional theory splitted from string theory: brane theory. From Nordstr\"{o}m \citep{nord} and Kaluza-Klein \citep{kalu,klein,klein1} initial five-dimensional scenarios, with one circular, compactified dimension and the constraint that the field must not depend on the extra-dimension, where developed three main classes of extra-dimensions \citep{jo,ruba}:
\begin{enumerate}
\item {\it Large extra-dimensions} \citep{anton1,arkani,arkani1,dick}\\
The Standard Model gauge and matter fields are confined to a three-dimensional brane that exists within a higher dimensional bulk. Only gravity can propagate into the compactified extra spatial dimensions.
\item {\it Warped extra-dimensions} \citep{rand,rand1}\\
The hierarchy between the Planck and the electroweak scales is generated by a large curvature of the extra-dimensions. The simplest framework in this scenario is is that of a five-dimensional Anti-de-Sitter ($AdS_{5}$) space, which is a space of constant negative curvature. In the compactified, finite extra-dimension just gravity propagates.
\item {\it TeV$^{-1}$-Sized extra-dimensions} \citep{anton}\\
By themselves, they do not allow for a reformulation of the hierarchy problem, but they may be incorporated into a larger structure in which this problem is solved. In these scenarios, the Standard Model fields are phenomenologically allowed to propagate in the bulk. This presents a wide variety of choices for model building:
\begin{itemize}
\item all, or only some, of the Standard Model gauge fields exist in the bulk;
\item the Higgs field may lie on the brane or in the bulk;
\item the Standard Model fermions may be confined to the brane or to specific locales in the extra-dimension. The phenomenological consequences of this scenario strongly depend on the location of the fermion fields.
\end{itemize}
\end{enumerate}

\section{Brane Formation by Particle Collision}

In principle, higher-dimensional objects can be formed also in particle collisions ({\it p}-branes). The standard picture is that when the colliding partons have a center-of-mass energy above some thresholds of order of the Planck mass ($\sim 10^{19}$ GeV) and the impact parameter less than the Schwarzschild radius, a Black Hole (type-0 brane) is formed and almost at rest in the center-of-mass frame. The Black Hole so produced will decay thermally and thus isotropically to that frame. In a universal extra-dimensions scenario, Kaluza-Klein (KK) states of quarks and gluons could be produced in colliders.

\begin{figure} [htp]
\centering
\leavevmode
\epsfxsize =7.0cm \epsfysize =3.0cm \epsfbox{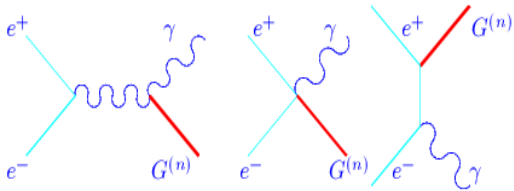}
\caption{Some interactions which can give rise to emission of real KK gravitons.}
\label{graviton1}
\end{figure}

In the large extra-dimensions scenario the first class of collider processes \citep{giu,mira,cheung} involves the real emission of KK graviton states in the scattering processes (\cite{jo}; see Fig. \ref{graviton1}): 

\[
e^{+}e^{-}\rightarrow \gamma + G^{(n)} \; {\rm or}\; Z + G^{(n)}
\]
\[
pp\rightarrow g + G^{(n)} \; ; \;
\bar{p}p\rightarrow g + G^{(n)}
\]
\[
Z\rightarrow f\bar{f} + G^{(n)}
\]

The produced graviton behaves as if it were massive, non-interacting, stable particle and thus appears as a missing energy in the detector.

At hadron colliders, for the interaction $q\bar{q}\rightarrow g + G^{(n)}$ is espected an emission process which to results in a monojet plus missing transverse energy signature.

\begin{figure} [htp]
\centering
\leavevmode
\epsfxsize =4.0cm \epsfysize =3.0cm \epsfbox{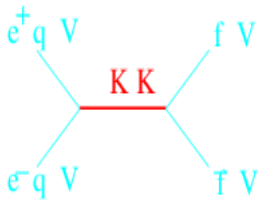}
\caption{Virtual KK graviton exchange possible processes.}
\label{graviton2}
\end{figure}

Deviations in cross sections and asymmetries in Standard Model processes as (see Fig. \ref{graviton2}):

\[
e^{+}e^{-}\rightarrow f\bar{f} \;{\rm or}\; \gamma\gamma \;{\rm or}\; ZZ \;{\rm or}\; WW
\]
\[
pp\rightarrow f\bar{f} \;{\rm or}\; \gamma\gamma
\]
\[
ep\rightarrow e + jet\; ,
\]
\noindent
or new production processes which are not present at three-level in the Standard Model as, for example $gg\rightarrow f\bar{f}$,
might be an evidence for a virtual graviton exchange \citep{giu,hew} but also putting limits on the extra-dimension size.

In the warped extra-dimension scenario, the principal expected collider signature should be the direct resonant production of the spin-2 states in the KK graviton tower \citep{davo,jo}.

In the $TeV^{-1}$ scaled extra-dimensions scenario, the masses of the excitation states in the gauge boson KK towers depend on where the Higgs boson is located. Precision electroweak data place strong constraints \citep{riz,masip,marciano} on the mass of the first gauge KK excitation.

For a detailed image of the physical processes in which are involved the KK states see Han et al. \citep{han}.

\section{Extra Dimensions Astrophysical and Cosmological Constraints}

Astrophysical and cosmological considerations impose strict constraints on some theories of extra-dimensions. A part of this constraints are resumed in table \ref{mass}.

\begin{table*}
\begin{center}
\begin{footnotesize}
\caption{Astrophysical and cosmological constraints on the fundamental mass scale of multi-dimensional gravity ($M\geq$ the table value in TeV; {\it d} is the number of extra-dimensions) \citep{jo}.}
\begin{tabular}{c c c c c c} \hline \hline
Constraints & {\it d} = 2 & {\it d} = 3 & {\it d} = 4 & {\it d} = 5 & References\\
\hline
Supernova Cooling & 30 & 2.5 & & & \citep{hanhart}\\
Cosmic Diffuse $\gamma$-Rays: & & & & & \\
1) {\bf Cosmic SNe} & 84 & 7 & & & \citep{hane}\\
2) $\nu \bar{\nu}$ Annihilation & 110 & 5 & & & \citep{hall}\\
3) Re-heating & 167 & 21 & 5 & 1.5 & \citep{hane1}\\
4) Neutron Star Halo & 450 & 30 & & & \citep{hane2}\\
Overclosure of Universe & 6.5/$\sqrt{h}$ & & & & \citep{hall}\\
Matter Dominated Early Universe & 85 & 7 & 1.5 & & \citep{fair}\\
Neutron Star Heat Excess & 1700 & 60 & 4 & 1 & \citep{hane2}\\
\hline
\end{tabular}
\label{mass}
\end{footnotesize}
\end{center}
\end{table*}

All the fundamental mass scale limits of multi-dimensional gravity from table \ref{mass} apply only for a situation where all extra-dimensions have the same compactification radius. In the general case the bounds could be less restrictive.

Ia Supernovae, Cosmic Microwave Background (CMB) and clusters data \citep{avelino}, also can impose restrictions on brane world scenarios through the values of the cosmological parameters. As well, cosmic ray experiments \citep{anchor1,feng,anchor,emparan} are imposing restriction through the absence of deeply neutrino penetrating showers, at rates above the Standard Model rate, initiated in the Earth's atmosphere by brane formation and decay ($\nu N\rightarrow$ {\it p}-brane)  \citep{kolbe,langanke,langanke1}.

\section{Size of Extra-Dimensions}

In the toroidal compactification assumption, in the case in which there is only one extra-dimension, for large extra-dimensions KK massive states there is an infinite-dimensional symmetry (the loop algebra on $S^{1}$) at the Lagrangian level \citep{dolan,aula,cho}, but it is broken by the vacuum configuration.

Han et al. paper's Introduction \citep{han} contains an important affirmation: "A torus compactification is perhaps not realistic, since the bulk fields which we are ignoring are potential sources of {\it n}-dimensional curvature, as are the branes themselves. However the torus has the great advantage of conceptual and calculation simplicity". So, just because its "simplicity", many of the authors, with some exceptions \citep{kalo}, in the field assume that the compactification is toroidal.

Also, in ADD \citep{arkani,arkani1} multi-dimensional scenario, the four-dimensional Planck scale is not a fundamental parameter. Rather, the mass scale of multi-dimensional gravity, which we will denote by {\it M}, is fundamental, as it is this latter scale that enters the full multi-dimensional action \citep{ruba}:

\begin{equation}
\label{eq_a}
S=-\; \frac{1}{16\pi\; G_{(D)}}\int{d^{D}X\; \sqrt{g^{(D)}}R^{(D)}}
\end{equation}
\noindent
where:

\[
G_{(D)}=\frac{1}{M^{D-2}}\equiv \frac{1}{M^{d+2}}
\]
\noindent
is the fundamental {\it D}-dimensional Newton's constant, {\it d} = {\it D} - 4 is the number of extra-dimensions, and $d^{D}X=d^{4}x\; d^{d}z$.

The long-distance four-dimensional gravity is mediated by the graviton zero mode whose wave function is assumed as {\bf homogeneous over extra-dimensions}. Hence, the four-dimensional effective action describing long-distance gravity is obtained from equation (\ref{eq_a}) by taking the metric {\bf to be independent of extra-coordinates {\it z}}. The integration over {\it z} becomes trivial, and the effective four-dimensional action:

\begin{equation}
\label{eq_b}
S_{eff}=\frac{V_{d}}{16\pi\; G_{(D)}}\int{d^{4}x\; \sqrt{g^{(4)}}R^{(4)}}
\end{equation}
\noindent
where $V_{d}\sim R^{d}$ is the volume of extra-dimensions. From eq. \ref{eq_b} we see that the four-dimensional Planck mass is equal to:

\begin{equation}
\label{eq_c}
M_{Pl}=R^{\frac{d}{2}}M^{\frac{d+2}{2}}
\end{equation}

Therefore, if {\it R} is large then {\it M} can be much smaller than $M_{Pl}$. If this scenario is to solve the hierarchy problem then $M\lesssim 10-100$ TeV \citep{hane,hane1}. Because of this requirement, {\it d} = 1 is excluded, $M\simeq 100$ TeV corresponding to $R\simeq 10^{8}$ cm.

But, if we will not consider the simplifying remarque that the graviton zero mode is homogeneous over extra-dimensions, but instead the graviton having an organized and structured internal distribution (as we will see further in D.E.U.S. model) and, even more, that metric and the Ricci scalar are {\bf not} independent of extra-dimensions, the integration over {\it z} in equation (\ref{eq_a}) will not be so trivial anymore. In consequence, in the four-dimensional effective action from equation (\ref{eq_b}) will enter supplementary terms, equation (\ref{eq_c}) becoming unrealistic and unable to describe in a right manner the extra-dimensions radius. The result is that the one-dimensional extra-dimension case won't be possible to be excluded in the above trivial manner.

In warped extra-dimensions scenarios (with emphasis on the Randall and Sundrum model) the effects of the branes on the bulk gravitational metric are taken into account, and are expressed in the action by the vacuum energy terms $V_{visible}$ and $V_{hidden}$ :

\begin{eqnarray}
\label{eq_d}
S=S_{gravity}+S_{visible}+S_{hidden}
\end{eqnarray}
\noindent
with:

\[
S_{gravity}=\int{d^{4}x}\int_{-\pi r}^{\pi r}{dz\; \sqrt{-g^{(5)}}[-\Lambda+2M^{3}R^{(5)}]}\; ,
\]
\[
S_{visible}=\int{d^{4}x\; \sqrt{-g_{visible}} [L_{visible}-V_{visible}]}\; ,
\]
\[
S_{hidden}=\int{d^{4}x\; \sqrt{-g_{hidden}} [L_{hidden}-V_{hidden}]}\; ,
\]
\noindent
where $\Lambda$ is the cosmological term, and the fifth-dimension bulk is located between $0 < z < z_{c}$, $z_{c}\equiv \pi r$ ({\it r} is the bulk radius). At each boundary plane it is assumed a $Z_{2}$ reflection symmetry. This results in a jump in the extrinsic curvature at these planes, yielding two domain branes of equal and opposite tension. One ends up with two three-brane rigidly located at the $S_{1}/Z_{2}$ orbifold fixed points in the {\it z} direction while extending in the $x^{\mu}$ directions. The solution to the five-dimensional Einstein's equations is required to respect four-dimensional Poincar\'{e} invariance in the $x^{\mu}$ directions, and is found to be:

\[
ds^{2}=e^{-2\sigma (z)}\eta_{\mu \nu}\; dx^{\mu}\; dz^{\nu}+dz^{2}\; ,
\]
\noindent
with $\sigma (z)=k\mid z\mid$. The $AdS_{5}$ curvature scale {\it k} is the single scale with fine-tunes between $\Lambda$, $V_{visible}$ and $V_{hidden}$, through $V_{hidden}=-V_{visible}=24M^{3}k$ and $\Lambda =-24M^{3}k^{2}$.

The relation between the Planck scale in four dimensions and the five-dimensional Planck scale {\it M} is found by integrating out the extra-dimension coordinate in the action:

\[
M_{Pl}^{2}=\frac{M^{3}}{k}\left(1-e^{-2\pi kr} \right) \; .
\]

Unlike the ADD scenario's equation (\ref{eq_c}), the two Planck scales in the Randall-Sundrum scenario are {\bf of the same order}.

\section{Acceleration and Branes in Supernova Shell}

\subsection{Massive Star Winds}

\begin{figure} [htp]
\centering
\leavevmode
\epsfxsize =7.0cm \epsfysize =7.0cm \epsfbox{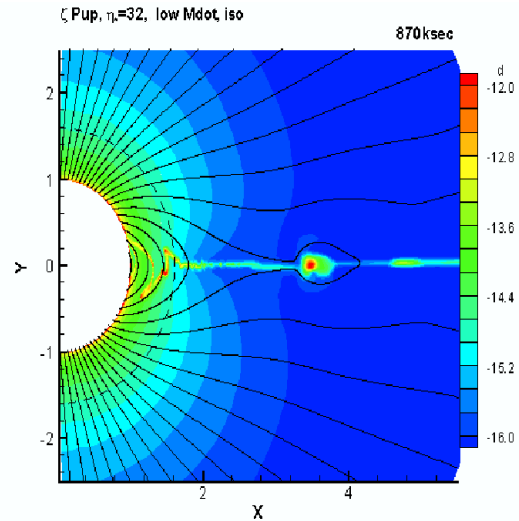}
\caption{Final state for $\zeta$ Pup ($M=50M_{\odot}$; $R=1.3\times 10^{12}$ cm; $\dot{M}=2.6\times 10^{-6}$ $M_{\odot}$/year; $v_{\infty}=2500$ km/s) with rotating wind, $\eta_{*}=32$ and $\omega \equiv \Omega/\Omega_{c}=1/2$simulation \citep{ud3}. The isothermal models with $\sqrt{10}\leq \eta_{*}\leq 10$ and no rotation show a similar final state behavior}
\label{wind}
\end{figure}

Into the paper \cite{pop} we see that for massive stars with strong winds ($30M_{\odot}\leq M\leq 50M_{\odot}$) the radiative acceleration plays an important role in the phase space dispersion of different wind particle species before injection in the supernova shock. Taking into account that for a massive star the degree to which the wind is influenced by the magnetic field can be described by just one dimensionless parameter: the wind magnetic confinement parameter, $\eta_{*}$ \citep{ud1,ud2}, meaning the ratio between the magnetic energy density and the kinetic energy density. ud-Doula \& Owocki simulations (see Fig. \ref{wind}) show that, for a standard O star, with $\eta_{*}\leq 1$ the magnetic field is fully opened by wind outflow, but that for $\eta_{*}=0.1$ it can enhance the matter density near the magnetic equator. For stronger confinement, $\eta_{*}>1$, the magnetic field remains closed over limited range of latitude and high above the equatorial surface, but eventually is opened into nearly radial configuration at large radii. Within the closed loops, the flow is channeled toward loop tops into perpendicular shock collisions. Within the open field region, the equatorial channeling leads to oblique shocks at the magnetic equator. Their assumption of isothermal hot-star wind is valid for the dense winds with the characteristic wind column densities much bigger than $7 \times 10^{21}$ cm$^{-2}$. For less dense winds with the columnar densities much smaller than $7 \times 10^{21}$ cm$^{-2}$ they used the adiabatic approximation. The simulation \citep{ud1,ud2} shows that, for $\eta_{*} < 1$ and identical stellar wind parameters, the solutions of isothermal and adiabatic models are similar. A different behavior can be seen for  $\eta_{*}\geq 1$, for which the adiabatic models exhibit an greatly extended magnetically confined region compared to their isothermal counterparts. These regions are filled in with very hot (up to $10^{8}$ K) low density gas that is in hydrostatic equilibrium. This is in contrast to isothermal confined region where hydrostatic atmospheres are not possible due to relatively low sound speed, and stellar gravity pull of the compressed, stagnated material within closed loops into an infall back onto the stellar surface through quite complex flow patterns. For the isothermal cases with zero rotation, unlike in the adiabatic models, there is a {\bf compressed dense, slowly outflowing equatorial disk} (see Fig. \ref{wind}). Instead, {\bf the region above the magnetic closed loops is very low density}.

We must accentuate the fact that, because for reaching an $\eta_{*}\sim1$ we will need an magnetic field of around 300 gauss for $\zeta$ Pup, a typical O-star with 50$M_{\odot}$, while for the solar wind we have an $\eta_{*}\sim40$. So, not just because of the different acceleration mechanism, we can not account for the same topology of the magnetic field and for its effects on the wind particles for the Sun or for an O-type star, especially on the same phase space distribution.

It must be emphasized that the Luminous Blue Variables (LBVs) giant outbursts may occur when these massive stars approach their Eddington limits. When this happens, they must reach a point where the centrifugal force and the radiative acceleration cancel out gravity at equator ($\Omega$-limit). When stars are close to the $\Omega$-limit, {\bf highly non-spherical mass loss} should occur. This suggests a scenario where {\bf a slow and very dense wind, strongly confined to the equatorial plane}, is followed by a fast and almost spherical wind \citep{langer}. With the assumption that the star is very close to the $\Omega$-limit Langer et al. \citep{langer} models produced gas distributions that strongly resemble Homunculus, the hourglass-shaped inner part of the highly structured circumstellar nebula.

The above models will consist our basic assumption for pre-supernova ejecta as a bipolar, dense, equatorial outflow.

\subsection{Shock Acceleration and Spallation of Cosmic Rays}

As was proposed by Biermann \citep{bier} and in Rachen \& Biermann \citep{rachen}, the cosmic rays observed events can be traced back to three sources \citep{wiebel}: 

\begin{quote}
\begin{enumerate}
\item Supernova explosions into the interstellar medium, or ISM-SNe. This component produces mostly hydrogen and the observed energetic electrons up to about 30 GeV, and dominates the {\it all particle} flux up to about $10^{4}$ GeV.
\item Supernova explosion into predecessor stellar wind, or wind-SNe. This component produces the observed electrons above 30 GeV, helium and most heavier elements already from GeV particle energies. Due to a reduction in acceleration efficiency at a particular
rigidity the slope of the spectrum increases, thus producing the knee feature. The component extends ultimately to several EeV. Since the winds of massive stars are enriched late in their life, this component shows a heavy element abundance which is strongly increased over that of the ISM.
\item Powerful radio galaxies produce a contribution which dominates beyond about 3 EeV, and consist mostly of hydrogen and helium, with only little addition of heavy elements. At this energy the interaction with the microwave background cuts off the contribution from distant extragalactic sources, the Greisen-Zatsepin-Kuzmin (GZK) cutoff. There are a small number of events which appear to be beyond this energy, and whether they fit into such a picture is open at present.
\end{enumerate}
\end{quote}

Various tests for these models were performed in Biermann \& Cassinelli \citep{bier1}, Biermann \& Strom \citep{bier2}, Stanev et al. \citep{stanev} etc.

The Cosmic Rays spectrum at GeV energies is close to $E^{-2.75}$, and at higher energies close to $E^{-2.65}$ below the "knee" at $\approx 5\cdot 10^{15}$ eV, from where the spectrum turns down to about $E^{-3.1}$, to flatten out again near $3\cdot 10^{18}$ eV (the "ankle"). The chemical composition is roughly similar to that of the interstellar medium, with reduced hydrogen and helium relative to silicon, and with some general enhancement of elements of low first ionization potential as we find in solar energetic particles.

The theory which claims to account for the origin of cosmic rays below $3\cdot 10^{18}$ eV traces them to the shock waves caused by supernovae exploding either into the interstellar medium (cosmic rays with energies bellow the "knee"), either into the predecessor stellar wind (cosmic rays with energies above the "knee" but bellow the "ankle") \citep{popescu1}.

Into the picture of a nonspherical wind as above, we will expect that, when the supernova shock will hit the material surrounding the star it will accelerate particles to cosmic ray energies through first order Fermi acceleration mechanism.
At the magnetic equator of our massive star, due to the dense wind presence, it will be possible to accelerate particles to $\sim 10^{18}$ eV. This acceleration limit will decrease dramatically from the equator to the polar regions because there practically the shock propagates in the ISM as it moves further away from the stellar surface. So, due to the time spent in shock (considered spherical), the maximal energy, which we will denote {\it E}, to which the particle will reach will be $\sim 10^{18}$ only in a narrow region around the equatorial plane, decreasing to around $\sim 10^{15}$ in the polar region. At low energies the isoenergetical  surfaces for the shock accelerated particles will follow closely the isodensity surfaces (see Fig. \ref{wind}). This situation changes for energies $10^{15} < E \leq 10^{18}$ eV, where the isoenergetical surfaces are following the isodensity profiles to around $\leq 30^{\circ}$ in stellar latitude after which are cut off. The resulting geometry are {\bf catenoid-like surfaces} centered on the star and having the stellar magnetic axis as the rotation symmetry axis.

The spallation is one of the processes that has one of the most important roles in the modification of the abundance of cosmic rays in their transportation through ISM \citep{reeves,silb,silb1}.
The basic features of target fragmentation, sometimes called "spallation", are very well understood: heavy fragments arise from peripheral collisions of heavy ions or relativistic protons with the target nucleus \citep{tsao,silb2,sihver}. These so called "spectators" of the reaction are excited primary fragments which then decay into the final fragments by a sequence of evaporation steps.

The spectrum of the residual nuclei seems to be determined to large extent, but not fully, by the evaporation process.

The difference between the spallation in high-energy physics in accelerators and in astrophysics is that, in the first case, the targets are the nuclei and the projectiles are the high-energy protons and in the second case the targets are the ISM protons (in first
approximation) and the projectiles are the nuclei that form the cosmic radiation. So everything depends just on which particle is put into the reference frame.

About 10\% by number of the interstellar gas is helium. Hence, about 20\% of cosmic ray generated nuclear spallation nuclear products are formed in nucleus-helium interactions. In the analysis of cosmic ray interactions with atmospheric nuclei (mainly nitrogen and oxygen) an accurate knowledge of nucleus-nucleus interactions becomes essential \citep{tsao}.

High-energy protons cause many different nuclear reactions and, in principle, all these nuclear processes have to be taken into account for a proper description of the total process. Possible outgoing particles are, for example, all light particles, gamma rays, {\bf neutrinos}, and (above incident energies of 150 MeV) pions. Furthermore, high-energy fission may occur and outgoing particles
from the proton bombardment stage further reactions. Clearly, a wide spectrum of reaction products will be formed.

The spallation happens when the supernova shock travels through the stellar wind and then hits the surrounding molecular shell of dense gas being responsible for the production of Li, Be and B in supernova shell. 

As well, the spallation with the ISM's protons or with the supernova predecessor star wind particles generates as secondary particles pions (but also kaons and muons) which decay afterwards into muons and neutrinos secondary particles \citep{gaisser}:

$$
\pi^{\pm}\rightarrow \mu^{\pm}+\nu_{\mu}(\bar{\nu}_{\mu})
$$
$$
\mu^{\pm}\rightarrow e^{\pm}+\nu_{e}(\bar{\nu}_{e})+\bar{\nu}_{\mu}(\nu_{\mu})\; ,
$$
\noindent
with a similar chain for charged kaons. Also:

$$
p+A\rightarrow \pi^{0}+B
$$
$$
\pi ^{0}\rightarrow \gamma \gamma \; .
$$

\section{Concluding Remarks}

An essential feature of the gravitational interaction is that at center-of-mass energies well above the fundamental scale, the gravitational coupling grows so large that graviton exchange dominates over all other interactions.

As said before, cosmic rays can probe the existence and the scale \citep{alvarez,feng,anchor,emparan} of extra-dimensions by observing the showers generated in Earth's atmosphere by brane decay. Black Holes decay thermally, according to the number of degrees of freedom available, and so their decays are mainly hadronic \citep{gidd,dimo}. For example, the non-perturbative Black Hole formation and decay by Hawking evaporation gives rise to hard jets and leptons, with a characteristic ratio of hadronic to leptonic activity of roughly 5:1. The high energy Black Hole cross section grows with energy at a rate determined by the dimensionality and {\bf geometry} of the extra-dimensions \citep{gidd}.

In this way, moving these physical arguments from Earth's atmosphere in the expanding supernova shell, even that the secondary neutrinos produced in spallation are in MeV energy range, if the fundamental scale is 1 TeV, their scattering on the $10^{18}$ eV accelerated particles may give rise to gravitons ($\sqrt{s}\sim 10^{5}$ GeV). Because the isoenergetical surfaces for these energies are catenoid-like surfaces, also the branes will form on a catenoid-like shaped surface, around the stellar remnant.

\begin{figure}
\centering
\leavevmode
\epsfxsize =7.0cm \epsfysize =7.0cm \epsfbox{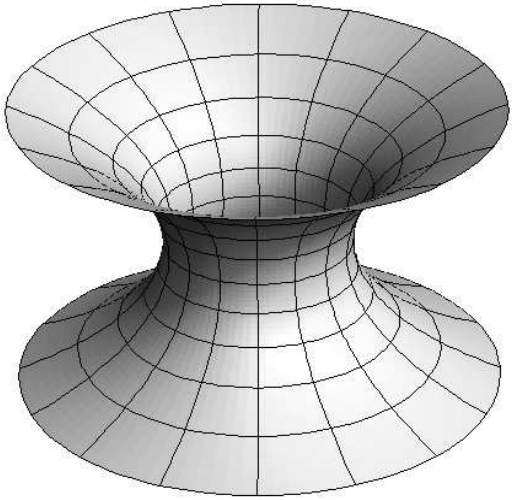}
\caption{Catenoid}
\label{catenoid}
\end{figure}

\begin{acknowledgements}
I want to express all my gratitude to Valeriu Tudose whose comments regarding this paper's clarity where very useful to me and, I belive, also to the reader.
\end{acknowledgements}

\end{document}